\documentclass[11pt]{article}
\usepackage{amsbsy}
\usepackage{amssymb,amsmath}
\usepackage{graphicx,latexsym,longtable}

\begin{document}

\title{Optimal Threshold Control by the Robots of Web Search Engines with Obsolescence of Documents}

\author{Konstantin Avrachenkov\thanks{INRIA Sophia Antipolis, France, K.Avrachenkov@sophia.inria.fr},
Alexander Dudin\thanks{Belarusian State University, Belarus, dudin@bsu.by},
Valentina Klimenok\thanks{Belarusian State University, Belarus, klimenok@bsu.by},\\
Philippe Nain\thanks{INRIA Sophia Antipolis, France, nain@sophia.inria.fr},
Olga Semenova\thanks{Belarusian State University, Belarus, olgasmnv@tut.by}
}


\date{}

\maketitle

\begin{abstract}
A typical Web Search Engine consists of three principal parts: crawling engine,
indexing engine, and searching engine. The present work aims to optimize the
performance of the crawling engine. The crawling engine finds new Web pages and
updates Web pages existing in the database of the Web Search Engine. The crawling
engine has several robots collecting information from the Internet. We first
calculate various performance measures of the system (e.g., probability of
arbitrary page loss due to the buffer overflow, probability of starvation of
the system, the average time waiting in the buffer). Intuitively, we would like
to avoid system starvation and at the same time to minimize the information loss.
We formulate the problem as a multi-criteria optimization problem and attributing
a weight to each criterion we solve it in the class of threshold policies.
We consider a very general Web page arrival process modeled by Batch Marked
Markov Arrival Process and a very general service time modeled by Phase-type
distribution. The model has been applied to the performance evaluation and
optimization of the crawler designed by INRIA Maestro team in the framework
of the RIAM INRIA-Canon research project.
\end{abstract}


%


\section{Introduction}
The problem of control by the robots (crawlers) that traverse the Web
and bring Web pages to the indexing engine that updates the data
base of a Web Search Engine is formulated and analyzed in
\cite{tlnc}. This problem is formulated in \cite{tlnc} as the
controlled queueing system. The system has a single server with the
exponential service time distribution, finite buffer of capacity
$K-1, \; K \ge 2.$ There are $N$ available robots and each of these
robots, when activated, brings pages to the server in a Poisson
stream at fixed rate.  These $N$ stationary Poisson processes are
mutually independent and independent of service times.

The number of active robots may be modified at any arrival or
departure epoch. When an arrival occurs, the incoming robot is
de-activated at once; the controller may then decide to keep it idle
or to activate it. When a departure occurs the controller may either
decide to activate one additional robot, if one is available, or to
do nothing (i.e. the number of active robots is left unchanged).

In \cite{tlnc}, the problem of finding a policy that minimizes a
weighted sum of the loss rate and starvation probability
(probability of the empty system) is considered. It is solved by
means of the tools of the Markov Decision Problems theory.

As the possible generalizations of the model, which are certainly
worthwhile analyzing, the following ones are mentioned in
\cite{tlnc}:
\begin{itemize}
\item[$\bullet$] More general input processes, e.g., a $MMPP$ ({\it
Markov Modulated Poisson Process}) should be considered so as to
reflect more accurately ``traveling times'' of robots in the network;
\item[$\bullet$] Because of the obsolescence of stored documents
issue, the waiting time should be bounded, even if the buffer size is
effectively infinite;
\item[$\bullet$] Other cost functions could be investigated, for
instance, cost functions including response times.
\end{itemize}

In this paper, we made all the mentioned and some further
generalizations.

We assume that, under the fixed number of currently active robots,
the arrival process is of the $BMAP$ type. The $BMAP$ is a more
general process comparing to the $MMPP$ and allows delivering of a
batch of Web pages to be indexed while the $MMPP$ assumes that the
pages are delivered one-by-one. It is very typical for a computer
system to operate in batch mode.

We assume that the service time distribution is of the $PH$ (Phase) 
type which is much more general comparing to the exponential distribution
assumed in \cite{tlnc}. The class of phase type distributions is
dense in the field of all positive-valued distributions and practically
we can deal with any real distribution \cite{ANO96}.

Since web pages can become obsolete, we bound the waiting time stochastically. 
Waiting time of each web page in a buffer is restricted by a random variable 
having $PH$ distribution identical and mutually independent for all Web pages.
The phase type distribution has been used to model obsolescence times
for instance in \cite{W87}.

We suppose that the cost function can have a more general form 
than in \cite{tlnc} and include the obsolescence probability and 
response time.

In the next section we formulate the model and optimization problem.
Section 3 contains the steady-state analysis of the
multi-dimensional Markov chain which defines dynamics of the system
under the fixed values of the parameters defining the strategy of
control. In Section 4, main performance measures of the system are
computed. In Section 5, the conditional sojourn time distributions
are calculated. In Section 6, the case of ordinary arrivals is touched in brief.
In Section 7, the theoretical results are illustrated by numerical examples.
In particular, the mathematical model is applied to the performance evaluation
and optimization of the robot designed by INRIA Maestro team in the
framework of the RIAM INRIA-Canon research project.
Section 8 concludes the paper.

\section{Mathematical Model}

We consider a single server system with the finite buffer of
capacity $K-1, \; K \ge 2.$ So, the total number of Web pages which
can stay in the system is restricted by the number $K.$ Web pages
are served by a server in order of their arrivals.

Service times of Web pages are independent identically distributed
random variables having $PH$ distribution with irreducible
representation $({\boldsymbol \beta}, S).$ It means the following.
Service of a Web page is defined as a time until the continuous-time
Markov chain $m_t,\; t \ge 0,$ having the states $(1,\dots,M)$ as
the transient and state $0$ as absorbing one reaches the absorbing
state. An initial state of the chain is selected in a random way,
according to the probability distribution defined by the row-vector
$({\boldsymbol \beta},0)$, where ${\boldsymbol \beta}$ is the
stochastic row vector of dimension $M$. Transitions of the Markov
chain $m_t,\;t \geq 0$, are described by the generator $ \left(
\begin{array}{cc}
 S&S_0\cr 0&0
\end{array}
\right)$
where the matrix $S$ is a sub-generator and the column
vector $S_0$ is defined by $S_0=-S{\bf e}_M\;$ and has all
non-negative and at least one positive components, ${\bf e}_M$ is
the column vector of dimension $M$ consisting of all 1's.
 The average service time $b_1$ is given by $b_1={\boldsymbol
\beta}(-S)^{-1}{\bf e}_M.$
For more details about the $PH$ type distribution, its properties,
special cases and applications see \cite{n81,pp}. 

Web pages can be delivered into the system by $N$ available robots.
The number of active robots varies in the set $\{1,\dots,N\}.$ We
assume that the process of Web pages delivering under $l,\;
l=\overline{1,N},$ active robots is described as follows. Let
$\nu_t, \; t \ge 0,$ be an irreducible continuous time Markov chain
having finite state space $\{0,1,\dots,W\}$. Sojourn time of the
chain $\nu_t, \, t\ \geq 0$, in the state $\nu$ has exponential
distribution with a parameter $\lambda^{(l)}_\nu$. After this time
expires, with probability $p^{(l)}_0(\nu,\nu')$ the chain jumps into
the state $\nu'$ without generation of Web pages and with
probability $p^{(l)}_k(\nu,\nu') $ the chain jumps into the state
$\nu'$ and a batch consisting of $k$ Web pages  is generated, $k
\geq 1$.  The introduced probabilities satisfy conditions:
$$
p^{(l)}_0(\nu,\nu)=0,\,\; \sum\limits_{k=1}^\infty
\sum\limits_{\nu'=0}^W p_k^{(l)}(\nu,\nu')+\sum\limits_{\nu'=0}^W
p_0^{(l)}(\nu,\nu')=1,\;\;
 \nu=\overline{0,W},\; l=\overline{1,N}.
 $$

The parameters defining this flow are kept in the square matrices
${\mathcal D}_k^{(l)},\,k\geq 0,\;l=\overline{1,N},$ of size $\bar
W=W+1$ defined by their entries:
$$({\mathcal D}^{(l)}_0)_{\nu,\nu}=-\lambda^{(l)}_\nu, \, ({\mathcal D}^{(l)}_0)_{\nu,\nu'}=\lambda^{(l)}_\nu
p_0(\nu,\nu'),\eqno(1)$$$$({\mathcal
D}_k^{(l)})_{\nu,\nu'}=\lambda^{(l)}_\nu p_k^{(l)}(\nu,\nu'),\,
\nu,\nu'=\overline{0,W},\, k\geq 1,\, l=\overline{1,N}.$$ Denote
$$
{\mathcal D}^{(l)}(z)=\sum\limits_{k=0}^{\infty} {\mathcal
D}^{(l)}_k z^k,\;|z| \le 1. $$

The matrix ${\mathcal D}^{(l)}(1)$ is the infinitesimal generator of
the process ${\nu_t}, t \geq 0,$ under the fixed number $l$ of
active robots. The stationary distribution vector ${\boldsymbol
\theta}^{(l)}$ of this process satisfies the equations
   $ {\boldsymbol \theta}^{(l)} {\mathcal D}^{(l)}(1)= {\boldsymbol 0},  {\boldsymbol \theta}^{(l)}{\bf e}=1.$
 Here and in the sequel, $ {\boldsymbol 0}$ is the zero row vector.
 The average intensity $\lambda^{(l)}$ (fundamental rate) of the $BMAP$ under the fixed number $l$ of
active robots is defined by
$$\lambda^{(l)}= {\boldsymbol \theta}^{(l)} \frac{d{\mathcal D}^{(l)}(z)}{d z}|_{z=1}{\bf e},$$
 and the intensity $\lambda^{(l)}_g$ of group arrivals is defined
 by $$\lambda^{(l)}_g= {\boldsymbol \theta}^{(l)}(-{\mathcal D}^{(l)}_0){\bf
 e}.$$
 The variance $v^{(l)}$ of  intervals between group arrivals is calculated as
$$v^{(l)} =2({\lambda^{(l)}_{g}})^{-1} {\boldsymbol
\theta}^{(l)}(-{\mathcal D}^{(l)}_0)^{-1}{\bf
e}-({\lambda^{(l)}_{g}})^{-2},$$ while the correlation coefficient
$c^{(l)}_{cor}$ of intervals between successive group arrivals is
given by
$$c^{(l)}_{cor}=\frac{1}{v(\lambda^{(l)}_{g})^{2}}\biggl(\lambda^{(l)}_{g}
{\boldsymbol \theta}^{(l)}(-{\mathcal D}^{(l)}_0)^{-1}({\mathcal
D}(1)^{(l)}-{\mathcal D}^{(l)}_0)(-{\mathcal D}^{(l)}_0)^{-1}{\bf
e}- 1\biggr).$$

The introduced representation of the arrival process via the matrices
${\mathcal D}_k^{(l)},\,k\geq 0,\;l=\overline{1,N},$ unifies several
possible interpretations of the process of Web pages delivered by
the fixed number of active robots:
\begin{itemize}
\item[1.] The processes of Web pages delivering by all robots are
independent $BMAP$ processes. Let the process of Web pages
delivering by the $l$th robot be the $BMAP$ which is governed by the
continuous time Markov chain $\nu_t^{(l)}, \; t \ge 0,$  having
finite state space $\{0,1,\dots,W_l\}$ and defined by the matrices
$D_k^{(l)},\,k\geq 0,\;$ of size $\bar W_l=W_l+1.$  See \cite{l} for
more details about the $BMAP$, its properties and special cases. We
denote $D^{(l)}(z)=\sum\limits_{k=0}^{\infty} D_k^{(l)} z^k,\;|z|
\le 1.$

Let us assume that the robots are arranged in such a way that the
first robot is always active, then, when a queue decreases, the
second robot can be activated, etc, the $N$th robot is the most rare
activated.

The matrices  ${\mathcal D}_k^{(l)},\,k\geq 0,\;l=\overline{1,N},$
of size $\bar W = \prod\limits_{k=1}^{N} \bar W_k$ defined by
formulae (1) are expressed via the matrices $D_k^{(l)},\;k\geq 0,\;$
of size $\bar W_l,\;l=\overline{1,N},$ describing the $BMAP$s in the
following way:
$$
{\mathcal D}^{(l)}_0=D_0^{(1)}\oplus D_0^{(2)}\oplus \dots \oplus
D_0^{(l)}\otimes D^{(l+1)}(1)\otimes \dots \otimes D^{(N)}(1),
$$
$$
{\mathcal D}^{(l)}_k=D_k^{(1)}\oplus D_k^{(2)}\oplus \dots \oplus
D_k^{(l)}\otimes I_{\bar W_{l+1}}\otimes \dots \otimes I_{\bar
W_{N}},\;\; k \ge 1.$$ Here $\otimes$ and $\oplus$ denote Kronecker
product and sum of matrices correspondingly (see, e.g.,
\cite{grah}), $I_L$ denotes identity matrix of size $L.$ If the size
of the matrix is clear from context the suffix can be omitted.
$I_0\stackrel{def}{=}1.$

\item[2.] The common process of Web pages delivered by all robots 
together is the $BMAP$ process
directed by the   continuous time Markov chain $\nu_t, \; t \ge 0,$
having finite state space $\{0,1,\dots,W\}$ and defined by the
matrices $D_k,\,k\geq 0,\;$ of size $\bar W.$ Some set of thinning
probabilities $q_1,\dots,q_N,\;$ $0< q_1 < \dots < q_N=1,\;$ is
fixed. When $l$ robots are active, procedure of thinning the $BMAP$
process with the thinning probability $q_l$ is applied. It means
that an arbitrary arriving batch is accepted with probability $q_l$
and is rejected with the complimentary probability $1-q_l.$ We
denote $D(z)=\sum\limits_{k=0}^{\infty} D_k z^k,\;|z| \le 1.$

The matrices  ${\mathcal D}_k^{(l)},\,k\geq 0,\;l=\overline{1,N},$
defined by formulae (1) are expressed via the matrices $D_k,\,k\geq
0,\;$  describing the common $BMAP$ and via the thinning probability
$q_l$ in the following way:
$$
{\mathcal D}^{(l)}_0=D_0q_l + D(1)(1-q_l),\;\; {\mathcal
D}^{(l)}_k=D_k q_l, \; k \ge 1.$$

\item[3.] Let  the process of Web pages delivering by robots is described by
a $BMMAP.$ The $BMMAP$ is directed   by continuous time Markov chain
$\nu_t, \; t \ge 0,$ having finite state space $\{0,1,\dots,W\}$.
Sojourn time of the chain $\nu_t, \, t\ \geq 0$, in the state $\nu$
has exponential distribution with a parameter $\lambda_\nu$. After
this time expires, with probability $p_0(\nu,\nu')$ the chain jumps
into the state $\nu'$ without generation of Web pages and with
probability $p^{(l)}_k(\nu,\nu') $ the chain jumps into the state
$\nu'$ and a batch consisting of $k,\;$$k \geq 1$ Web pages  are
delivered by the $l$th robot.  The introduced probabilities satisfy
conditions:
$$p_0(\nu,\nu)=0,\,\,
\sum\limits_{l=1}^N\sum\limits_{k=1}^\infty \sum\limits_{\nu'=0}^W
p_k^{(l)}(\nu,\nu')+\sum\limits_{\nu'=0}^W p_0(\nu,\nu')=1,\;\;
 \nu=\overline{0,W}.$$

The matrices  ${\mathcal D}_k^{(l)},\,k\geq 0,\;l=\overline{1,N},$
defined by formulae (1) are expressed via the matrices
$D_0,\;D^{(l)}_k,\,k\geq 1,\;l=\overline{1,N},$ formed by the
probabilities $p_0(\nu,\nu)$ and $p_k^{(l)}(\nu,\nu')$ in the
following way:
$$
{\mathcal
D}^{(l)}_0=D_0+\sum\limits_{j=1}^{\infty}\sum\limits_{m=l+1}^{N}D^{(m)}_j,
\;{\mathcal D}^{(l)}_k=\sum\limits_{m=1}^{l}D^{(m)}_k,\;k \ge 1.
$$
\item[4.] The  process of Web pages delivery by all robots is the $BMAP$ process
directed by the   continuous time Markov chain $\nu_t, \; t \ge 0,$
having finite state space $\{0,1,\dots,W\}$ and defined by the
transition intensity matrices ${\mathcal D}_k^{(l)},\,k\geq
0,\;l=\overline{1,N},$ depending on the number of active robots.
\end{itemize}
Interpretation 3 seems be the most attractive because it assumes
that the work of the robots can be dependent, which is quite
realistic. Because the total amount of Web servers from which new
pages should be brought is more or less constant, reduction of the
number of active robots causes the increase of field in Internet,
which is scrawled by each robot, and corresponding change of travel
time. So, the $BMMAP$ looks to be the most realistic model of Web pages
delivery.

If a batch of delivered Web pages meets free server one
Web page starts the service immediately while the rest moves to the
buffer. If the server is busy at an arrival epoch, all Web pages of
the batch are placed into the buffer if there is enough free space
in the buffer. If the number of free places in the buffer is less
than the number of Web pages in the batch, the corresponding number
of Web pages is lost. It means that we consider so called partial
admission strategy. The alternative strategies of complete rejection
or complete admission can be investigated in analogous way.

For each Web page placed into the buffer, the waiting time is
 restricted by the random
variable (so called obsolescence time) having $PH$ distribution with
irreducible representation $({\boldsymbol \gamma}, \Gamma).$ It
means the following. Available waiting time of the $i$th Web page in
the buffer is defined as a time   until the continuous-time Markov
chain $r_t^{(i)},\; t \ge 0,$ having the states $(1,\dots,R)$ as the
transient and state $0$ as absorbing one, reaches the absorbing
state. Transition of this process into the absorbing state means
that this Web page gets out of date (obsolescence or dashout
occurs). An initial state of the chain is selected in a random way,
according to the probability distribution defined by the row-vector
$({\boldsymbol \gamma},0)$, where ${\boldsymbol \gamma}$ is the
stochastic row vector of dimension $R$. Transitions of the Markov
chain $r_t^{(i)},\;t \geq 0$, are described by the generator $
\left(
\begin{array}{cc}
 \Gamma& \Gamma_0\cr 0&0
\end{array}
\right)$
 where the matrix $\Gamma$ is sub-generator and the column
vector $\Gamma_0$ is defined by $\Gamma_0=-\Gamma{\bf e}.$
 The average  time until obsolescence $g_1$ is given by $g_1={\boldsymbol
\gamma}(-\Gamma)^{-1}{\bf e}.$

If the obsolescence time expires before a Web page is picked-up from
the buffer to the server, it is assumed  that this Web page
immediately leaves the buffer and is lost. The obsolescence times of
different Web pages are independent of each other and identically
distributed. It is worth to note that the analysis presented below
could be drastically simplified if we suggest that the obsolescence
time is exponentially distributed. However, this suggestion rarely
holds true in the real world systems because this suggestion means
that, with high probability, information obsoletes very quickly.

Reasonable class of strategies of control by robots is the class of
the threshold strategies defined as follows. Integers
$j_1,\dots,j_{N-1}$ are fixed such as $-1=j_0<j_1<\dots<
j_{N-1}<j_N=K.$ If the number $i$ of Web pages in the system
satisfies inequality $j_{r-1}+1 \le i \le j_r,$ then $N-r+1$ robots
are active and $r-1$ robots are de-activated, $r=\overline{1,N}.$

Note that the described threshold strategies are popular in
literature in controlled queues, see, e.g., \cite{aod,c,d,dk,t}.
For some systems, it is proven that the optimal strategy in the
class of all Markovian strategies belongs to the class of threshold
strategies. For some other systems such a result is not proven, but
the optimal strategy is sought in the class of threshold strategies.
Advantage of such strategies is their intuitive justification and
relative simplicity of implementation in real-life systems.
Numerical examples presented in the paper \cite{tlnc} for a partial
case of our model confirm that the threshold strategies are optimal
in the class of all Markovian strategies, although authors cannot
prove this fact. Our system is much more complicated and we also 
cannot prove optimality of the optimal threshold strategy in wider
classes of strategies. We just try to find an optimal threshold
strategy and believe that it is optimal or sub-optimal in wider
classes as well.

We also mention that the description of the given above threshold strategy
suits only for the case when $N \le K.$ While the numerical examples
presented in \cite{tlnc} address, e.g., the case $N=16$ and $K=5$.
However, if we look at the optimal strategy given by Figure 2 in
\cite{tlnc}, we see that the optimal number of the active robots
varies between 4 and 14 with 4 switching points where two or three
robots are activated or de-activated.  Because, in contrast to
\cite{tlnc}, we do not assume that each active robot generates a
stationary Poisson process of arrivals at a fixed rate but we assume
that each robot generates a batch Markovian arrival process,
we see that our strategy suits for the case $N
>K$ as well. We could achieve this formally by allowing non-strict inequalities:
 $-1=j_0 \le j_1\le \dots\le j_{N-1}\le j_N=K$ in fixing the
 thresholds.
 So, speaking below about a robot we may think about a virtual robot
 as a group of several
 available real robots which are activated and de-activated
 simultaneously.

We will solve the problem of choosing the optimal threshold
strategy. The cost function is assumed to be of the following form:
$$
J=\lambda (c_{loss} P_{loss} + c_{obs} P_{obs}) + a \bar V^{(1)} +
c_{rob} N_{act} + c_{star} P_{star} \eqno(2)
$$
where $\lambda$ is the mean number of pages delivered into the
system by robots during a unit of time (fundamental rate of the
arrival process), $P_{loss}$ is probability of arbitrary page loss
due to the buffer overflow, $P_{obs}$ is probability of arbitrary
page obsolescence during waiting in a queue, $P_{star}$ is
probability of starvation of the system,  $\bar V^{(1)}$ is the
response time (average sojourn time of Web pages which are not lost
or deleted due to obsolescence), $N_{act}$ is the average number 
of active robots, $c_{loss}, c_{obs}, a, c_{rob},c_{star}$ are the
corresponding non-negative cost coefficients. The values of the cost
coefficients can be set up by experts in the domain. Alternatively,
the cost coefficients can be viewed as Lagrange multipliers in the
constrained problem and can be found from the dual problem 
formulation.

It is clear that if the average number of active robots is
increasing then the three first summands in (2) are increasing while
the last summand, charge for starvation of the system, is
decreasing. The last charge is important because starvation means
that the indexing machine is idle and so freshness of the
data base suffers.

The problem of minimization of the cost criterion (2) is not
trivial. To solve this problem, we will use so called direct
approach. To this end, we will calculate the stationary distribution
of the system state under an arbitrary fixed set of  thresholds
$(j_1,\dots,j_{N-1}).$ It will allow us to calculate the main
performance measures of the system and the value $J$ of the cost
criterion as a function $J(j_1,\dots,j_{N-1}).$ Problem of finding
the optimal set $(j^{*}_1,\dots,j^{*}_{N-1})$ is then easy solved on
computer, e.g., by enumeration.

\section{Stationary distribution of the number of Web pages in the system}

Let some set of  thresholds $(j_1,\dots,j_{N-1})$ be fixed. We are
interested in the stationary distribution of the process $i_t,\; t
\ge 0,$ where $i_t$ is the number of Web pages in the system
at the epoch $t,\;$$i_t=\overline{0,K}.$ This process is
non-Markovian. To investigate this process, we will consider the
following multi-dimensional continuous time Markov chain
$$
\xi_t=\{i_t, \nu_t, m_t, r_t^{(1)},\dots,r_t^{(i_t-1)}\}, \; t \ge
0,
$$
where $\nu_t$ is the state of the directing process of arrival
process at epoch $t,\;$ $\nu_t=\overline{0,W};$ $m_t$ is the state
of the process which directs a service at epoch $t,\;$
$m_t=\overline{1,M};$ $r_t^{(i)}$ is the state of the process, which
directs a obsolescence of the $i$th Web page in a queue at epoch
$t,\;$ $r_t^{(i)}=\overline{1,R},\; i=\overline{0,K-1}.$

We assume that  the Web pages in the buffer are numerated in order
of their arrival into the system. If a batch of Web pages arrives
  to the system, the  accepted Web pages are numerated in a uniform
random manner. When a Web page is picked up to the service or is
deleted from the queue because its admissible waiting time expires,
the rest of Web pages is immediately  enumerated correspondingly.

Denote
$$
p(0,\nu)= \lim\limits_{t\to\infty}P\{i_t=0,\nu_t=\nu\},
$$
$$
p(1,\nu,m)= \lim\limits_{t\to\infty}P\{i_t=1,\nu_t=\nu, m_t=m
\},\eqno(3)
$$
$$
p(i,\nu,m,r_1,\ldots,r_{i-1})=$$$$
\lim\limits_{t\to\infty}P\{i_t=i,\nu_t=\nu, m_t=m,
r_t^{(1)}=r_{1},\ldots,r_t^{(i-1)}=r_{i-1}\},\;\; i \ge 2.$$ Because
the state space of the the Markov chain $\xi_t, t\geq 0,$ is finite
and due to assumption about irreducibility of the processes defining
arrival, service and obsolescence processes, limits (3) exist.

Enumerate the states of the Markov chain $\xi_t, t\geq 0,$ in the
lexicographic order and form the probability row vectors ${\bf p}_i,
i=\overline{0,K},$ of probabilities corresponding to the state $i$
of the first component of the process $\xi_t, t\geq 0$. Denote also
${\bf p}=({\bf p}_0,\dots,{\bf p}_K)$.

Let $Q$ be the infinitesimal generator of the Markov chain $\xi_t,
t\geq 0,$ and $Q_{i,i'}$ be the block of the generator $Q$
consisting of enumerated in lexicographic order intensities of
transition of the Markov chain $\xi_t, t\geq 0,$ from the states
with the value $i$ of the component $i_t$ to the states with the
value $i'$ of this component, $i,i' \ge 0.$ Dimension of the block
$Q_{i,i'}$ is defined by $(\bar W M^{a_i}R^{b_i})\times(\bar W
M^{a_{i'}}R^{b_{i'}})$ where
$a_l=\min\{l,1\},\;$$b_l=\max\{(l-1),0\},\;l=\overline{0,K}.$

 {\it Lemma.} ~Non-zero blocks $Q_{i,i'}$ of the infinitesimal generator $Q$ of the Markov chain $\xi_t, t\geq
 0,$ are defined by
$$
Q_{0,j}=\left \{ \begin{array}{lr} {\mathcal D}^{(N)}_0,\;\; j=0,
\cr {\mathcal D}^{(N)}_j\otimes {\boldsymbol \beta} \otimes
{\boldsymbol \gamma}^{\otimes (j-1)}, \;\; j=\overline{1,K-1}, \cr
({\mathcal
 D}^{(N)}(1) - \sum\limits_{r=0}^{K-1} {\mathcal D}^{(N)}_r)\otimes {\boldsymbol \beta} \otimes {\boldsymbol
\gamma}^{\otimes (K-1)}, \;\; j=K, \cr
\end{array} \right.
$$
$$
Q_{i,i-1}=\left \{ \begin{array}{lr} I_{\bar W}\otimes S_0,\;\; i=1,
\cr I_{\bar W}\otimes {\mathcal B}_{i-1}, \; i > 1, \cr
\end{array} \right.
$$
$$
Q_{i,i}=\left \{ \begin{array}{lr} {\mathcal D}^{(\chi(i))}_0 \oplus
{\mathcal A}_{i-1},\;\; i=\overline{1,K-1}, \cr {\mathcal
D}^{(\chi(i))}(1)\oplus {\mathcal A}_{i-1},\;\; i=K, \cr
\end{array} \right.
$$
$$
Q_{i,i+l}=\left \{ \begin{array}{lr} {\mathcal D}^{(\chi(i))}_l
\otimes I_{M R^{i-1}} \otimes{\boldsymbol \gamma}^{\otimes l},\;\;
l=\overline{1,K-i-1},\;  i< K-1, \cr ({\mathcal D}^{(\chi(i))}(1) -
\sum\limits_{r=0}^{K-i-1} {\mathcal D}^{(\chi(i))}_r)\otimes I_{M
R^{i-1}}\otimes {\boldsymbol \gamma}^{\otimes l},\;\;  l=K-i, \;\;i<
K,\cr
\end{array} \right.
$$
$$i  > 0.$$
 Here

$$
{\mathcal B}_i = S_0{\boldsymbol \beta}\otimes {\bf e}_R \otimes
I_{R^{i-1}} +I_M \otimes \Gamma_0^{\oplus i},\; {\mathcal
A}_i=S\oplus \Gamma^{\oplus i},\; i=\overline{0,K-1},
$$
$${\boldsymbol \gamma}^{\otimes
l}\stackrel{def}{=}\underbrace{{\boldsymbol
\gamma}\otimes\ldots\otimes{\boldsymbol \gamma}}_l,\; \Gamma^{\oplus
l}\stackrel{def}{=}\underbrace{\Gamma\oplus\ldots\oplus \Gamma}_l,\;
\Gamma_0^{\oplus
l}\stackrel{def}{=}\underbrace{\Gamma_0\oplus\ldots\oplus
\Gamma_0}_l,\quad l\geq 1,
$$
and the value $\chi(i),$ which corresponds to the number of active
robots when $i$ Web pages stay in the system, is calculated by
$\chi(i)=N-\psi(i)$ where the value $\psi(i),
\;\psi(i)=\overline{0,N-1},$ is defined by relations
$$
j_{\psi(i)}+1 \le i \le j_{\psi(i)+1}, \; i=\overline{0,K}.
$$

 Proof of the lemma is implemented by analyzing the probabilities of
transition of the multi-dimensional Markov chain $\xi_t, t\geq 0,$
under the fixed set of  thresholds $(j_1,\dots,j_{N-1})$ during an
interval of infinitesimal length. It is rather transparent and is
omitted.

It is well-known that the vector ${\bf p}$ of stationary
probabilities satisfies the following equilibrium equations
$$
{\bf p} Q = {\bf 0},\;\; {\bf p}{\bf e}=1. \eqno(4)
$$

Dimension of the vector ${\bf p}$ is equal to $\bar W
(1+M\frac{R^{K-1}-1}{R-1}).$ It can be rather high. For instance, if
$\bar W=M=R=2$ this dimension is equal to $2^{K+1}-2.$ For $K=10$
this equals to 2046. So, direct solution of equations (4) by "brute
force" can be very time and computer memory demanding.

 Effective and
numerically stable algorithm for computing the blocks ${\bf p}_i,
i=\overline{0,K},$ of the vector ${\bf p}$, which exploits a
structure of generator $Q$ and is presented in \cite{kkod}, consists
of the following steps:
\begin{itemize}
\item[1.] Compute the matrices $G_i,\,i=\overline{0,K-1},$ from the
backward recursion
$$G_i=(-Q_{i+1,i+1}-\sum\limits_{l=1}^{K-i-1}Q_{i+1,i+1+l}G_{i+l}G_{i+l-1}\dots G_{i+1})^{-1}
 Q_{i+1,i},\;$$
 $$i=K-1,K-2,\dots,0.$$
\item[2.] Compute the matrices $\overline{Q}_{i,l}$ from the
backward recursions
$$\overline{Q}_{i,K}=Q_{i,K}, i=\overline{0,K},
$$
$$\overline{Q}_{i,l}=Q_{i,l}+\overline{Q}_{i,l+1}G_l, i=\overline{0,l}, l=K-1, K-2,\dots,0.$$
\item[3.]
Compute the matrices $F_l,\,l=\overline{0,K},$ by recursion
$$F_0=I,\;F_l=\sum\limits_{i=0}^{l-1}F_i \bar Q_{i,l} (-\bar Q_{l,l})^{-1},
l=\overline{1,K}.
$$
\item[4.] Compute the vector ${\bf p}_0$ as the unique solution to the following system
of linear algebraic equations:
$${\bf p}_0\overline{Q}_{0,0}={\bf 0},\;{\bf p}_0\sum\limits_{l=0}^K F_l {\bf e}=1.$$
\item[5.] Compute the vectors ${\bf p}_i,\; i=\overline{1,K},$ by
formula
$${\bf p}_i={\bf p}_0 F_i,\,i=\overline{1,K}. $$
\end{itemize}
Thus, the problem of computing the stationary distribution ${\bf
p}_i, \;i=\overline{0,K},$ of the considered queueing system under
the arbitrary fixed set of the   thresholds $(j_1,\dots,j_{N-1})$
can be considered solved.

\section{Main Performance Measures of the System}

As the main performance measures of the system we consider the
values $ \lambda,\; P_{loss},\; P_{obs},\; P_{star},\; N_{act}$
which appear in cost criterion (2).

Calculation of the values $P_{star},\;  N_{act}$ is the easiest.

{\it Theorem 1.} Probability $P_{star}$ of the system starvation
(idle state of the indexing machine) is calculated by
$$
P_{star}={\bf p}_0{\bf e}_{\bar W}. \eqno(5)
$$
Average number $N_{act}$ of active robots is calculated by
$$
N_{act}= \sum\limits_{n=1}^{N} n \varphi_n, \eqno(6)
$$
where probability $\varphi_n$ that $n$ robots are active at an
arbitrary epoch is computed by
$$
\varphi_n = \sum\limits_{i=j_{N-n}+1}^{j_{N+1-n}} {\bf p}_i {\bf
e}_{\bar W M^{a_i}R^{b_i}}, n=\overline{1,N}.\eqno(7)
$$
{\it Theorem 2.} Probability $P_{loss}$ of an arbitrary Web page
loss due to the buffer overflow is calculated by
$$
P_{loss}=1-\frac{1}{\lambda} \sum\limits_{n=0}^{N-1}
\sum\limits_{i=j_{n}+1}^{j_{n+1}}{\bf p}_i \sum\limits_{k=0}^{K-i}
(k-K+i)({\mathcal D}^{(N-n)}_k \otimes I_{M^{a_i}R^{b_i}}){\bf e},
\eqno(8)
$$
where the average intensity $\lambda$ of the input flow is computed
by
$$
\lambda=\sum\limits_{n=0}^{N-1}
\sum\limits_{i=j_{n}+1}^{j_{n+1}}{\bf p}_i
\sum\limits_{k=1}^{\infty} k({\mathcal D}^{(N-n)}_k\otimes
I_{M^{a_i}R^{b_i}}){\bf e}. \eqno(9)
$$
Proof. It follows from the formula of total probability that
$$
P_{loss}=1-\sum\limits_{i=0}^{K}\sum\limits_{k=1}^{\infty} P^{(k)}_i
P_k \Phi^{(i,k)},\eqno(10)
$$
where $P^{(k)}_i$ is probability that there are $i$ Web pages in the
system at an epoch of arrival of a batch consisting of $k$
Web pages, $P_k $ is probability that arbitrary Web page arrives in
the batch consisting of $k$ Web pages, $\Phi^{(i,k)}$ is probability
that the arbitrary Web page will be accepted into the system
conditional that it arrives in the batch consisting of $k$ Web pages
and $i$ Web pages present in the system at the epoch of arrival.

It can be shown that the listed probabilities are calculated by
$$
P^{(k)}_i =\frac{{\bf p}_i ({\mathcal D}^{(N-n)}_k \otimes
I_{M^{a_i}R^{b_i}}){\bf e}}{\sum\limits_{m=0}^{N-1}
\sum\limits_{r=j_{m}+1}^{j_{m+1}}{\bf p}_r ({\mathcal D}^{(N-m)}_k
\otimes I_{M^{a_r}R^{b_r}}){\bf e}},\eqno(11)
$$
$$
j_{n}+1 \le i \le j_{n+1}, \;n=\overline{1,N-1},
$$
$$
P_k=\frac{\sum\limits_{n=0}^{N-1}
\sum\limits_{i=j_{n}+1}^{j_{n+1}}{\bf p}_i k({\mathcal D}^{(N-n)}_k
\otimes I_{M^{a_i}R^{b_i}}){\bf e}}{\sum\limits_{n=0}^{N-1}
\sum\limits_{i=j_{n}+1}^{j_{n+1}}{\bf p}_i
\sum\limits_{l=1}^{\infty} l ({\mathcal D}^{(N-n)}_l \otimes
I_{M^{a_i}R^{b_i}}){\bf e}},\eqno(12)
$$
$$
\Phi^{(i,k)}=\left \{ \begin{array}{lr} 1,\;\; k \le K-i, \cr
\frac{K-i}{k}, \;\; k > K-i, \cr
\end{array} \right. \eqno(13)
$$$$
i=\overline{0,K}, k \ge 1.
$$
By substituting expressions (11) - (13) into formula (10) we get
formulae (8), (9). The theorem is proven.

{\it Theorem 3.} Probability $P_{obs}$ of an arbitrary Web page
obsolescence is calculated by
$$
P_{obs}= \frac{1}{\lambda} \sum\limits_{i=2}^{K}{\bf p}_i (I_{\bar
W} \otimes I_M \otimes \Gamma_0^{\oplus (i-1)}){\bf e}. \eqno(14)
$$
Probability $P_{success}$ of an arbitrary Web page successful
service in the system is calculated by
$$
P_{success}= \frac{1}{\lambda} \sum\limits_{i=1}^{K}{\bf p}_i
(I_{\bar W} \otimes S_0 \otimes I_{R^{i-1}}){\bf e}. \eqno(15)
$$
The statement of the theorem is clear because the right hand side of
(14) represents the ratio of the obsolescence rate and arrival rate
into the system. The right hand side of (15) represents the ratio of
the  rate of  successfully  served in the system Web pages and
arrival rate into the system.

\section{Sojourn Time Distribution}

Let $V(x),\;$ $V_1(x),\;$ and $V_2(x)$ be distribution functions of
sojourn time of an arbitrary Web page in the system under study, an
arbitrary Web page, which will get successful service, and arbitrary
Web page, which will be deleted from the system due to its
obsolescence, and $v(u),\; v^{(1)}(u)$ and $v^{(2)}(u) $ be the
corresponding Laplace-Stieltjes transforms:
$$
v(u)=\int\limits_{0}^{\infty} e^{-u x} d V(x),\;\;
v^{(k)}(u)=\int\limits_{0}^{\infty} e^{-u x} d V_k(x),\; k=1,2,\;\;
Re\;u\;> 0.
$$

{\it Theorem 4.}  Laplace-Stieltjes transforms $v(u),\; v^{(1)}(u)$
and $v^{(2)}(u) $ are calculated by
$$
v(u)=v^{(1)}(u) P_{success}+ v^{(2)}(u)P_{obs} + P_{loss}, \eqno(16)
$$
 $$v^{(1)}(u)= \frac{1}{ P_{success}}\sum\limits_{i=0}^{K-1} \sum\limits_{k=1}^{\infty} {\bf
 p}_{i,k}
 {\bf v}^{(1)}_{i,k}(u), \eqno(17)
$$
$$
 v^{(2)}(u)= \frac{1}{ P_{obs}}\sum\limits_{i=0}^{K-1} \sum\limits_{k=1}^{\infty} {\bf p}_{i,k}
 {\bf v}^{(2)}_{i,k}(u), \eqno(18)
$$
where
$$
{\bf p}_{i,k}=\frac{1}{\lambda}{\bf p}_i k
 ({\mathcal D}^{(\chi(i))}_k{\bf e}_{\bar W}  \otimes
I_{M^{a_i}R^{b_i}}),
$$
the column vectors ${\bf v}^{(m)}_{i,k}(u), i=\overline{0,K-1},\; k
\ge 1, \; m=1,2,$ are computed by
$$
{\bf v}^{(m)}_{i,k}(u)=\sum\limits_{l=1}^{\min \{k,K-i\}}{\mathcal
C}_{i,l} \frac{1}{k}{\bf v}^{(m)}_{i+l-1}(u),\;m=1,2,\eqno(19)
$$
$$
{\mathcal C}_{i,l}={\boldsymbol \beta}^{\delta_{i,0}}\otimes
I_{M^{a_i}R^{b_i}}\otimes {\boldsymbol \gamma}^{\otimes
(l-\delta_{i,0})},$$
 the vectors ${\bf
v}^{(m)}_{i}(u),\; i=\overline{0,K-1},\; m=1,2,$ are computed
recursively by
$$
{\bf v}^{(1)}_{0}(u)=(uI-S)^{-1}S_0,\eqno(20)
$$
$$
{\bf v}^{(1)}_{i}(u)= (uI -{\mathcal A}_{i} )^{-1}\hat {\mathcal
B}_{i-1} {\bf v}^{(1)}_{i-1}(u),\; i =\overline{1,K-1},
$$
$$
{\bf v}^{(2)}_{0}(u)=0,\; {\bf v}^{(2)}_{i}(u)=
\sum\limits_{l=0}^{i-1} \prod\limits_{k=0}^{l-1} (uI - {\mathcal
A}_{i-k})^{-1}\hat {\mathcal B}_{i-k-1} \times
$$$$\times(uI - {\mathcal A}_{i-l})^{-1}(I_M \otimes
I_{R^{i-l-1}}\otimes \Gamma_0) {\bf e},\; i =\overline{1,K-1},
\eqno(21)
$$
where
$$
\hat {\mathcal B}_i = {\mathcal B}_i \otimes I_R, \;
i=\overline{0,K-1}.
$$
$\delta_{i,l}$ is Kronecker delta: $\delta_{i,l}=\left
\{\begin{array}{rc} 1, \; i=l, \cr 0, i \ne l. \cr
\end{array}\right .$

Proof. To prove the theorem, we use the method of collective marks
(method of catastrophes). We interpret the parameter $u$  as an
intensity of some imaginary stationary Poisson flow of catastrophes.
So, $v(u)$ has the meaning of probability that no catastrophe
arrives during the sojourn time of an arbitrary Web page,
$v^{(1)}(u)$  is probability that no catastrophe arrives during
the sojourn time of an arbitrary Web page conditional  that this
Web page will get service successfully, and $v^{(2)}(u)$ is
probability that no catastrophe arrives during the sojourn time
of an arbitrary Web page conditional  that this Web page will be
deleted from the system due to its obsolescence.

So, formula (16) evidently stems from the formula of total
probability.

It is obvious that sojourn time of the arbitrary (tagged) Web page
does not depend on the arrival process after the tagged Web page
arrival epoch. Thus,  in analysis we can ignore transitions of the
directing process of the arrival process after the epoch of the
tagged Web page arrival.

Formulae (17) and (18) also follow from the formula of total
probability. Here row vector ${\bf  p}_{i,k}$ defines probability
of an arbitrary Web page arrival at the moment when there are $i$
Web pages in the system in a batch of size $k$ and probability
distribution of the directing processes of service and obsolescence
at this moment. Column vector ${\bf v}^{(m)}_{i,k}(u)$ defines
probability of no catastrophe arrival during the conditional
sojourn time of a tagged Web page who arrives at the moment when
there are $i$ Web pages in the system in a batch of size $k$ under
the fixed value of the directing processes of service and
obsolescence at the arrival epoch. For $m=1$ condition is that the
tagged Web page will get service successfully. For $m=2$ condition
is that the tagged Web page will be deleted from the system due to
its obsolescence.

Relation (19) is based again on the formula of total probability.
The matrix $ {\mathcal C}_{i,l}$ defines probability distribution of
installation, upon the arrival epoch of a tagged Web page,  of the
initial states of the directing process of service (if $i=0$, i.e.,
the system was empty at the arrival epoch) and of the directing
processes of obsolescence of the Web pages, which arrive at the same
batch as the tagged Web page and are placed in a buffer before the
tagged Web page, and  of this Web page. Recall that we assume that
 if the batch consists of $k$ Web pages then the
tagged Web page will be the $l$th in the batch, $l=\overline{1,k},$
with probability $\frac{1}{k},\; k \ge 1.$

The vector $ {\bf v}^{(m)}_{i}(u)$ defines probability of no
catastrophe arrival during the conditional sojourn time of a tagged
Web page who sees $i$ Web pages in the system before him in a queue
 and the corresponding states of the
directing processes of service and obsolescence after the moment of
arrival.

Recurrent formulae (20) for the vector Laplace-Stieltjes transforms
$ {\bf v}^{(1)}_{i}(u)$ is clear if we take into account that: (i) $
{\bf v}^{(1)}_{0}(u)$ is the vector Laplace-Stieltjes transform of
the service time distribution, (ii) $(uI -{\mathcal A}_{i} )^{-1}$
defines probability that no catastrophe arrives during the time
interval between the epoch of a tagged Web page arrival, at which
the number of Web pages in a queue before the tagged Web page is
equal to $i$, and the epoch when the number of Web pages in a queue
before the tagged Web page is decreased to $i-1$, (iii) the matrix
$\hat {\mathcal B}_i$ defines transition of the directing processes
of service and obsolescence of the Web pages at the epoch of
decreasing.

Formulae (21) for the vector Laplace-Stieltjes transforms $ {\bf
v}^{(2)}_{i}(u)$ takes into account reasonings (ii),(iii) presented
above as well as consideration that no catastrophe should arrive
until the obsolescence moment of a tagged Web page and that $l,\;
l=\overline{0, i-1},$ Web pages can depart from the system until the
obsolescence moment. The theorem is proven.

{\it Corollary 1.}  Average sojourn time (response time) $\bar V$ of
an arbitrary Web page, average sojourn time  $\bar V^{(1)} $ of an
arbitrary Web page who will get successful service, and average
sojourn time  $\bar V^{(2)} $ of an arbitrary Web page who will be
deleted from the system due to its obsolescence are computed by
$$
\bar V= \bar V^{(1)} P_{success}+ \bar V^{(2)} P_{obs}, \eqno(22)
$$
$$\bar V^{(1)}=  \frac{1}{ P_{success}}\sum\limits_{i=0}^{K-1} \sum\limits_{k=1}^{\infty} {\bf
 p}_{i,k}
 {\bf w}^{(1)}_{i,k}, \eqno(23)
$$
$$
 \bar V^{(2)} =  \frac{1}{ P_{obs}}\sum\limits_{i=0}^{K-1} \sum\limits_{k=1}^{\infty}  {\bf p}_{i,k}
 {\bf w}^{(2)}_{i,k}, \eqno(24)
$$
where the column vectors ${\bf w}^{(m)}_{i,k}=-\frac{ d {\bf
v}^{(m)}_{i,k}(u)}{d u}|_{u=0},\; i=\overline{0,K-1},\; k \ge 1, \;
m=1,2,$ are computed by
$$
{\bf w}^{(m)}_{i,k}=\sum\limits_{l=1}^{\min \{k,K-i\}}{\mathcal
C}_{i,l} \frac{1}{k}{\bf w}^{(m)}_{i+l-1},\;m=1,2,\eqno(25)
$$
 the vectors ${\bf w}^{(m)}_{i}, \;i=\overline{0,K},\; m=1,2,$ are
computed  by
$$
{\bf w}^{(1)}_{0}=-S^{-1} {\bf e}_M,\;\; {\bf w}^{(1)}_{i} = -(
{\mathcal A}_{i})^{-1}\biggl[{\bf v}^{(1)}_{i}(0) + \hat {\mathcal
B}_{i-1} {\bf w}^{(1)}_{i-1} \biggr], i =\overline{1,K-1},
$$
$$
{\bf v}^{(1)}_{0}(0)={\bf e}_M,\; {\bf v}^{(1)}_{i}(0)= -( {\mathcal
A}_{i})^{-1}\hat {\mathcal B}_{i-1} {\bf v}^{(1)}_{i-1}(0),\; i
=\overline{1,K-1},
$$
$$
{\bf w}^{(2)}_{i}=- \sum\limits_{l=0}^{i-1} (-1)^{l} \biggl[
\sum\limits_{m=0}^{l-1} \prod\limits_{k=0}^{m-1} ({\mathcal
A}_{i-k})^{-1} \hat {\mathcal B}_{i-k-1} \times \biggr.
$$
$$
\times ({\mathcal A}_{i-m})^{-2} \hat{\mathcal B}_{i-m-1}
\prod\limits_{k=m+1}^{l-1} ({\mathcal A}_{i-k})^{-1} \hat {\mathcal
B}_{i-k-1} +
$$
  $$
  +\prod\limits_{k=0}^{l-1} ({\mathcal A}_{i-k})^{-1}
\hat{\mathcal B}_{i-k-1}({\mathcal A}_{i-l})^{-1}\biggr]({\mathcal
A}_{i-l})^{-1}(I_{M R^{i-l-1}} \otimes \Gamma_0){\bf e}_M, \;$$
$$ i =\overline{1,K-1}.
$$
Proof of corollary evidently follows from the well-known expression
for the mean value of a random variable via the derivative of the
Laplace-Stieltjes transform of its distribution function.

Note that expressions for higher order moments and variance on
sojourn time distribution can be also easily derived based on
equations (16)-(18).

\section{Case of the Ordinary Arrival Process}
Consider the special case when Web pages arrive not in batches, but
one-by-one. It means that ${\mathcal D}_k^{(l)}=0,\,k\geq
2,\;l=\overline{1,N}.$ In this case  the generator $Q$ of the Markov
chain $\xi_t, t\geq 0,$ is the three block diagonal matrix which in turn
means that this chain is a finite space
Quasi-Birth-and-Death-Process. Thus, in this case the algorithm for
solving equilibrium equations (4) for the vector ${\bf p}$ of
stationary probabilities and formulae for some performance measures
simplify. The algorithm has the form
\begin{itemize}
\item[1.] Compute the matrices $G_i,\,i=\overline{0,K-1},$ from the
backward recursion
$$G_i=[-(Q_{i+1,i+1}+Q_{i+1,i+2}G_{i+1})]^{-1}Q_{i+1,i},\;i=K-2,K-3,\dots,0,$$
with the terminal condition
$$G_{K-1}=-(Q_{K,K})^{-1}Q_{K,K-1}.$$
\item[2.] Compute the matrices $F_i,\,i=\overline{0,K},$ by recursion
$$F_0=I,\;\;F_i=F_{i-1}Q_{i-1,i}[-(Q_{i,i}+Q_{i,i+1}G_i)]^{-1},\,i=\overline{1,K}.$$
\item[3.] Compute the vector ${\bf p}_0$ as the unique solution to the following system
of linear algebraic equations:
$${\bf p}_0(Q_{0,0}+Q_{0,1}G_0)={\bf 0},\;{\bf p}_0\sum\limits_{l=0}^K F_l {\bf e}=1.$$
\item[4.] Compute the vectors ${\bf p}_i,\; i=\overline{1,K},$ by
formula
$${\bf p}_i={\bf p}_0 F_i,\,i=\overline{1,K}. $$
\end{itemize}
Formula for loss probability $P_{loss}$ is given by
$$
P_{loss}=\frac{1}{\lambda} {\bf p}_K ({\mathcal D}^{(1)}_1 \otimes
I_{M R^{K-1}}){\bf e},
$$
where
$$
\lambda=\sum\limits_{n=0}^{N-1}
\sum\limits_{i=j_{n}+1}^{j_{n+1}}{\bf p}_i ({\mathcal D}^{(N-n)}_1
\otimes I_{M^{a_i}R^{b_i}}){\bf e}.
$$
Laplace-Stieltjes transforms $ v^{(1)}(u)$ and $v^{(2)}(u) $ are
computed by
$$v^{(1)}(u)= \frac{1}{\lambda P_{success}}\sum\limits_{i=0}^{K-1} {\bf
 p}_{i}({\mathcal D}^{(\chi(i))}_1{\bf e}_{\bar W}  \otimes
I_{M^{a_i}R^{b_i}}){\mathcal C}_{i,1}
 {\bf v}^{(1)}_{i}(u),
$$
$$
 v^{(2)}(u)= \frac{1}{\lambda P_{obs}}\sum\limits_{i=0}^{K-1} {\bf p}_{i}
 ({\mathcal D}^{(\chi(i))}_1{\bf e}_{\bar W}  \otimes
I_{M^{a_i}R^{b_i}}){\mathcal C}_{i,1}
 {\bf v}^{(2)}_{i}(u).
$$
Average sojourn times   $\bar V^{(1)} $ and $\bar V^{(2)} $ are
computed by
$$\bar V^{(1)}=  \frac{1}{ \lambda P_{success}}\sum\limits_{i=0}^{K-1} {\bf
 p}_{i}({\mathcal D}^{(\chi(i))}_1{\bf e}_{\bar W}  \otimes
I_{M^{a_i}R^{b_i}}){\mathcal C}_{i,1}
 {\bf w}^{(1)}_{i},
$$
$$
 \bar V^{(2)} =  \frac{1}{ \lambda P_{obs}}\sum\limits_{i=0}^{K-1} {\bf p}_{i}
 ({\mathcal D}^{(\chi(i))}_1{\bf e}_{\bar W}  \otimes
I_{M^{a_i}R^{b_i}}){\mathcal C}_{i,1} {\bf w}^{(2)}_{i}.
$$
\section{Numerical example}

To demonstrate feasibility of the developed algorithms for
calculating the stationary state distribution of the system under
the fixed parameters of the control strategy and calculating the
optimal set of these parameters, let us consider numerical examples.
First we suppose that the system can have any number of active
robots between one and four at any time moment ($N=4$) and the
buffer capacity is equal to $5$ ($K=5$).

We assume that the arrival process is formed according to Model~4
from Section~2. Namely, when $r$ robots are activated the
$BMAP$-input is described by the matrices ${\mathcal D}_k^{(r)}$,
$k\geq 0$, $r=\overline{1,N}$, given by

\begin{gather*}
D_0^{(1)}=\begin{pmatrix}
  -10 & 2 \\
  0 & -0.5 \\
\end{pmatrix},\;
D_1^{(1)}=\begin{pmatrix}
  0.05 & 3 \\
  0.29 & 0.005 \\
\end{pmatrix},
D_2^{(1)}=\begin{pmatrix}
  0.05 & 4.9 \\
  0.2 & 0.005 \\
\end{pmatrix},\,\\
D_0^{(2)}=\begin{pmatrix}
  -5.65 & 1.1 \\
  0 & -0.85 \\
\end{pmatrix},
D_1^{(2)}=\begin{pmatrix}
  -0.01 & 2.5 \\
  0.25 & 0 \\
\end{pmatrix},\,D_2^{(2)}=\begin{pmatrix}
  -0.02 & 0.5 \\
  0.25 & 0 \\
\end{pmatrix},\\D_3^{(2)}=\begin{pmatrix}
  0.02 & 1.5 \\
  0.1 & 0.25 \\
\end{pmatrix},\,
D_0^{(3)}=\begin{pmatrix}
  -2.48 & 0.48 \\
  0.48 & 3.48 \\
\end{pmatrix},\\D_1^{(3)}=\begin{pmatrix}
  1.5 & 0 \\
  0 & 2.25 \\
\end{pmatrix},\,
D_2^{(3)}=\begin{pmatrix}
  0.5 & 0 \\
  0 & 0.75 \\
\end{pmatrix},\\D_0^{(4)}=\begin{pmatrix}
  -1.45 & 0.45 \\
  0.6 & -2.6 \\
\end{pmatrix},\,
D_1^{(4)}=\begin{pmatrix}
  0.25 & 0 \\
  0 & 0.5 \\
\end{pmatrix},\\
D_2^{(4)}=D_3^{(4)}=\begin{pmatrix}
  0 & 0 \\
  0 & 0 \\
\end{pmatrix},\,
D_4^{(4)}=\begin{pmatrix}
  0.75 & 0 \\
  0 & 1.5 \\
\end{pmatrix}.
\end{gather*}

The intensities $\lambda^{(r)}$ of the $BMAP$ when $r$ robots are
active, $r=\overline{1,N},$ are computed by
$$\lambda^{(1)}=1.28,\; \lambda^{(2)}=2.41,\;\lambda^{(3)}=3.125,\; \lambda^{(4)}=4.64.$$
The  coefficients of correlation $c_{cor}^{(r)}$ and the intensities
of  batches arrival $\lambda^{(r)}_g$ are the following:
$c_{cor}^{(1)}=-0.218$, $c_{cor}^{(2)}=-0.111$,
$c_{cor}^{(3)}=0.02$, $c_{cor}^{(4)}=0.035$,
$\lambda^{(1)}_g=0.853$, $\lambda^{(2)}_g=1.208$,
$\lambda^{(3)}_g=2.5$, $\lambda^{(4)}_g=1.43$.

Let the $PH$ distribution of service time be defined by the row
vector ${\boldsymbol \beta}=(0.4\;0.6)$ and sub-generator
$S=\begin{pmatrix}
  -3  & 1 \\
  2 & -3 \\
\end{pmatrix}.$  The mean service
time is equal to 0.657.

Let the $PH$ distribution of a Web page obsolescence time be
described by the row vector $\gamma=(0.3\; 0.7)$ and  sub-generator
$\Gamma=\begin{pmatrix}
  -0.6 & 0.4 \\
  0.1 & -0.3 \\
\end{pmatrix}.$  The mean time
until obsolescence is equal to 5.

Let the cost  coefficients be fixed by $c_{loss}=5$, $c_{obs}=10$,
$a=2$, $c_{rob}=20$, $c_{star}=300.$ We have chosen the cost coefficients
in this way to obtain commensurable optimal values in the optimal solution
and non-trivial optimal policy.

Note that in this example we fixed the cost coefficients based on
some heuristic reasonings or common sense. In general, the right
choice of the cost coefficients is the important and difficult task.
It requires good knowledge of the real world system, which is
described by the mathematical model under study, and clear
understanding what is the most undesirable for the concrete system
(loss of the delivered Web pages, obsolescence of a page, starvation
of the system, long waiting in the queue, keeping to many robots be
active, etc), what is less important. So, the help of experts is
required for the right choice of the cost coefficients. If such a
choice does not seem be possible, some alternative formulation of
optimization problem, e.g., multi-criteria problem or problem with
constraints may be considered. In the latter approach the cost
coefficients are Lagrange multipliers in the constrained problem
and can be found from the dual problem formulation. 

Let us find the optimal strategy of control by the system under the
fixed above values of the system parameters and the cost
coefficients. The thresholds $j_1,\dots,j_{N-1}$ in the problem
formulation are fixed such as $-1=j_0<j_1<\dots< j_{N-1}<j_N=K,$
i.e., the thresholds cannot coincide and the use of  all $N$ modes
of operation (the number of the mode is characterized by the number
of the active robots) is mandatory. It is intuitively clear that
actually it can happen that the optimal strategy does not need to
use some modes of operation at all. So, to find the optimal
strategy, we have to compare the values of the optimal values of the
cost criterion when all $N$ modes are used, when $N-1$ modes are
used while 1 mode is ignored, $\dots$, two modes are used while
$N-2$ modes are ignored, when only one mode is used (i.e., the
number of the active robots is not varied).

Let us denote by $C_r$ the value of the cost criterion when exactly
$r$ robots are always active, $r=\overline{1,4}.$ The values $C_r$,
$r=\overline{1,4},$ are given by $C_1=149.91$, $C_2=110.0$,
$C_3=89.405$, $C_4=130.312$. So, if there is no possibility to
control the number of robots, and one has to decide how many robots
should permanently work, the best choice is  to have permanently
three active robots.

Next we consider the threshold type strategies for controlling the
number of active robots. Table 1 contains the optimal value of the
cost criterion for various combinations of the used modes and the
optimal threshold strategy for each such a combination.
\begin{center}
Table 1: The Value of the Optimal Cost Criterion for various
threshold strategies
\begin{tabular}{|c|c|c|}
 \hline possible numbers of active robots & \begin{tabular}{c}
  Optimal \\
  thresholds \\
\end{tabular} & \begin{tabular}{c}
  Optimal value \\
of the cost \\
  criterion \\
\end{tabular}
\\\hline
1&--&149.91\\ 2&--&110.0\\3&--&89.40\\4&--&130.31\\
  2 or 1&2& 103.54\\
 \textbf{3 or 1} & \textbf{2} &\textbf{63.54}\\
  4 or 1&1&74.47\\
  3 or 2&2&76.21\\
  4 or 2&1 &86.13 \\
   4 or 3&0 &94.14 \\
     3 or 2 or 1&2,2 &63.54 \\
 4 or 2 or 1&1,2 &73.69 \\
  4 or 3 or 1& 0,2&80.50 \\
4 or 3 or 2&0,2 &80.50\\
 4 or 3 or 2 or 1&0,2,2 &67.52 \\
  \hline
  \end{tabular}
\end{center}

\vspace{4mm}

Let us explain the entries of the table. For instance, the
highlighted line corresponds to the control with one threshold at
two. When the queue length does not exceed two, there are three
active robots and when the queue length exceeds two, the number of
active robots decreases to one. As another example, the control
corresponding to the last line of the table has the following
structure: when the system is empty, four robots are active; when
the queue length is greater than zero and smaller than three, three
robots are active; when the queue length exceeds two, only one robot
is active.

As it is seen from Table 1, the optimal strategy assumes that only
 two among four available operation
modes (modes with one and three active robots) should be used. The
optimal value $C^*$ of the cost criterion is equal to 63.54. It is
evident that $C^*$ gives the relative profit $R$ more than 28\%
comparing to the case without control. The value of $R$ is computed
as
$$R=\left(1-\frac{C^*}{\min\{C_1,C_2,C_3,C_4\}}\right)*100\%.$$

The dependence of the cost criterion on the threshold when the
strategy of control  uses only two modes, for all possible
combinations of the modes, is shown on Figure 1.

\begin{figure}[htb]
 \centering \includegraphics[scale=0.6]{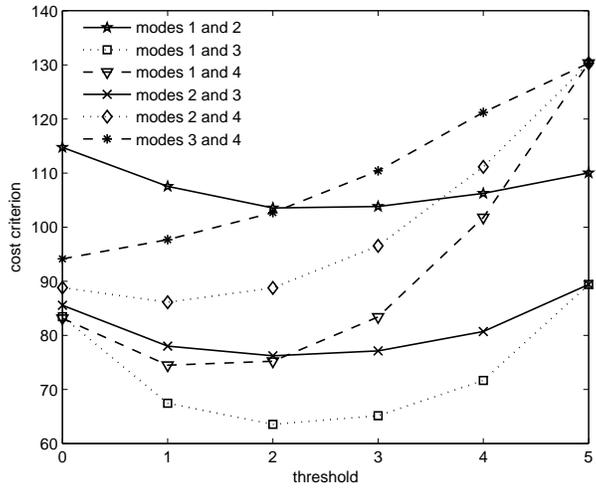}\\
  \caption{Dependence of the cost criterion on the threshold}
\end{figure}

The obtained results illustrate the necessity of the input control
and possibility to reduce the cost of the system operation by means
of the threshold type control.

Now let  us vary the buffer size $K$  from 1 to 100. Table 2
contains the optimal criterion value for the cases with and without
control ($C^*$ and $C_r$, $r=\overline{1,4}$), the optimal threshold
$j^*$ and the relative profit $R$ for different buffer capacity $K$.
Note that in all cases the optimal control only uses the modes with
one and three active robots. The results for $K>30$ are
approximately the same as for $K=30$.

\newpage
\begin{center}
Table 2: Variable Buffer Size $K$
\\ \vspace{2mm}
\begin{tabular}{|c|c|c|c|c|c|c|c|}
  \hline
  $K$ & $j^*$ & $C^*$ & $C_1$ & $C_2$ & $C_3$ & $C_4$ & $R$, \% \\
  \hline
  1 & 0 & 147.5& 244.7 & 233.4 & 187.2 & 258.8 & 21.0 \\
  2 & 1 & 96.8 & 199.2 & 174.0 & 128.8 & 194.4 & 24.8 \\
  3 & 2 & 79.1 & 172.6 & 140.3 & 105.4 & 160.0 & 24.9 \\
  4 & 2 & 68.3 & 158.1 & 121.7 & 94.7 & 140.6 & 27.85\\
  5 & 2 & 63.5 & 149.9 & 110.0 & 89.4 & 130.3 & 28.9 \\
  6 & 3 & 60.8 & 144.7 & 102.3 & 86.7 & 124.1 & 29.8 \\
  7 & 3 & 59.3 & 141.6 & 97.2 & 85.5 & 120.5 & 30.6 \\
  8 & 3 & 58.4 & 138.5 & 93.7 & 85.0 & 118.3 & 31.2 \\
  9 & 3 & 57.8 & 137.9 & 91.3 & 84.9 & 117.1 & 31.8\\
10 & 3  & 57.5 & 137.0 & 89.7 & 85.0 & 116.5 & 32.3 \\
20 & 3  & 57.2 & 137.0 & 86.3 & 87.2 & 120.0 & 33.7\\
30 & 3  & 57.2 & 137.0 & 86.3 & 87.4 & 123.6 & 33.7\\
  \hline
\end{tabular}\end{center}

\vspace{4mm}

Let the mean service time $b_1$ be varied by means of the matrix $S$
multiplication by the value $s$ which varies from 0.1 to 15. Table 3
contains the value $s$, the mean service time $b_1$, the optimal set
of possible numbers of active robots, the optimal value of the
threshold, the optimal cost criterion value and the value of the
cost criterion when only one of the modes is in use (when the number
of active robots is fixed), and the relative profit $R$. Note that
when $s$ is equal to or grater than 15, i.e. $b_1$ is less than
0.04, no dynamic control is required.

Now we vary the mean time $g_1$ until obsolescence by the same way
as it was done for the mean service time. Table 4 shows obtained
results. Note that in the optimal set of modes only one and three
active robots are used in all considered cases.

Consider the effect of the service time variation on the cost
criterion value given that the mean service time $b_1$ is constant
and equals to 0.657. Let the matrix $S$ of service time distribution
have the form
$$S=\begin{pmatrix}
  -\alpha_1 & 0 \\
  0 & -\alpha_2 \\
\end{pmatrix}$$
and vector ${\boldsymbol {\boldsymbol \beta}} = (0.9\; 0.1)$.

The variance of service time is calculated as
$$var_s = 2{\boldsymbol {\boldsymbol \beta}} S^{-2}{\mathbf{e}} - ({\boldsymbol {\boldsymbol \beta}} (-S)^{-1}\mathbf{e})^2.$$
To maintain the mean service time $b_1$, the entries $\alpha_1$ and
$\alpha_2$ of matrix $S$ must be related through the formula
$b_1={\boldsymbol {\boldsymbol \beta}}(-S)^{-1}{ \mathbf{e}}_M$ as
$$\alpha_2 = \frac{0.1\alpha_1}{b_1\alpha_1-0.9}.$$
Note that $\alpha_1$ should be grater than 1.369 to keep $\alpha_2$
positive.

Let us vary the value of $\alpha_1$ from 1.521 to 9. The service
time variation $var_s$ takes the values from 0.431 to 5.798. In the
case $\alpha_1=1.521$ service time distribution is exponential one.
In the optimal set of operation modes only one and three active
robots are used. Figure 2 shows dependence of the cost criterion
value on the threshold under different values of service time
variation ($\alpha_1\in\{1.521, 3,5,7,9\}$).

\begin{figure}[htb]
 \centering \includegraphics[scale=0.6]{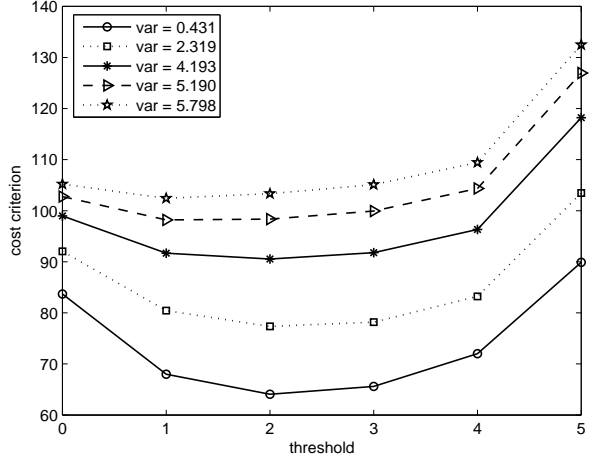}\\
  \caption{Dependence of the cost criterion on service time variation}
\end{figure}

But when $\alpha_1$ becomes larger than 4 ($var_s>3.415$), not the
modes with one and three active robots  but the modes with one and
four active robots are the optimal set of exploited modes. Figure 3
shows the dependence of the cost criterion when only optimal set of
modes is in use. In the figure, two lower curves correspond to the
modes with one and three active robots and other curves correspond
to the modes with one and four active robots.

\begin{figure}[htb]
 \centering \includegraphics[scale=0.6]{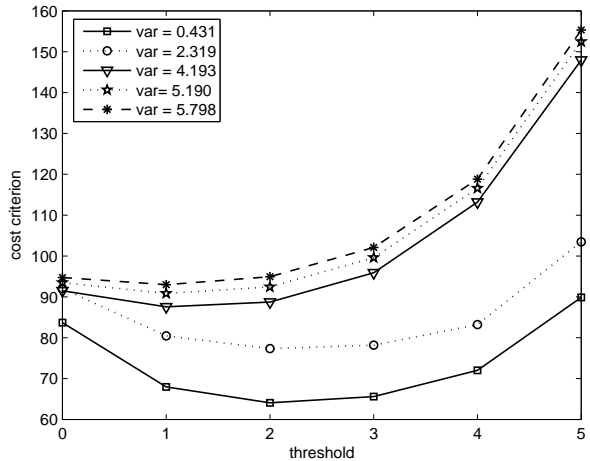}\\
  \caption{Dependence of the cost criterion on service time variation}
\end{figure}

Now let us vary the variance of time until obsolescence in the same
way. Let the matrix $G$ have the form
$$G=\begin{pmatrix}
  -\alpha_1 & 0 \\
  0 & -\alpha_2 \\
\end{pmatrix}$$
and $\gamma = (0.9\; 0.1).$ To maintain the average time until
obsolescence $g_1=5$, the value $\alpha_2$ is related to $\alpha_1$
as
$$\alpha_2 = \frac{0.1\alpha_1}{g_1\alpha_1-0.9}.$$
We vary $\alpha_1$ from 0.2 to 7. Thus the variation takes the
values from 25 to 449. Note that in the case $\alpha_1=0.2$, we get
the time until obsolescence distributed exponentially. The optimal
strategy consists of modes with one and three active robots in all
considered cases. Figure 4 shows dependence of the cost criterion
value on variation of time until obsolescence ($\alpha_1\in\{0.2,
0.5,1,3,7\}$). Note that as the variation grows the optimal
threshold decreases from 2 to 0.

\begin{figure}[htb]
 \centering \includegraphics[scale=0.6]{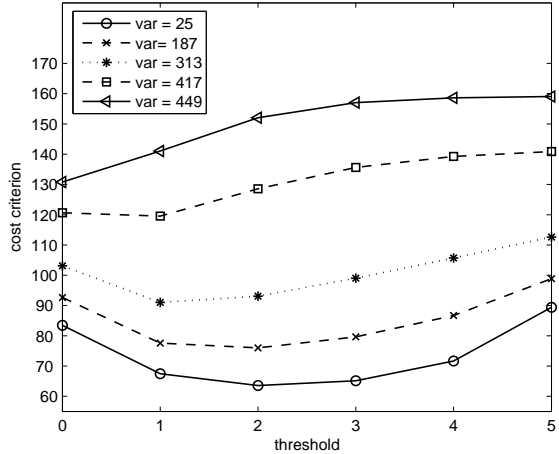}\\
  \caption{Dependence of the cost criterion on variation of time
until obsolescence}\label{fig:3.3}
\end{figure}

Now let us consider the example based on real data obtained from
the robot designed by INRIA Maestro team in the framework of RIAM
INRIA-Canon Research Project. Data about the information
delivery process by the robot to the data base was presented
in the form of the text file which contained more than 65 000
timestamps defining epochs of information delivery.

This data was
processed in the following way.
\begin{itemize}
\item[$\bullet$]
Inter-arrival times were computed.
\item[$\bullet$]
The obtained sample was censored: very long intervals, which
actually correspond to the periods when the crawling process was
stopped due to some reason, are deleted from the sample.
\item[$\bullet$]
The obtained sample was transformed into two samples in such a way
as the very short inter-arrival times were deleted from the initial
sample and the corresponding information about the number of
successively deleted intervals was placed into another sample. As
the result of these manipulations, we stated that the arrival
process is the batch arrival process. One sample defines the
intervals between the epochs of batches arrival, the second sample
defines the number of the information units in the corresponding
batch. By information units we mean either the principal part of
a Web page (e.g., Web page main html file) or its embedded resources
(e.g., image and audio files).
\item[$\bullet$]
Based on the second sample, distribution of the number of the
information units in a batch was computed as follows. The size of a
batch varies in the interval $[1,8]$ and probabilities $d_k$ that
the batch size is equal to $k=1,2,\dots,8$ are the following:
$d_1=0.040445227$, $d_2=0.18404232$, $d_3=0.26405114$,
$d_4=0.5057307$, $d_5=8.8163983\cdot 10^{-4}$, $d_6=7.7143486\cdot
10^{-4}$, $d_7=0.0018734847$, $d_8=0.00220405$. The mean batch size
is equal to
 3.263389905223716.
 \item[$\bullet$]
 Based on the first sample, we computed estimation of the mean value of an interval between
 arrivals of batches as 212.03,
estimation of the variance of such an interval as 51352.38 and the
estimations of lag-$k$ correlation of inter-arrival times for $k$
equal to $1,2,\dots,6$ are given by  0.622;  0.574; 0.555; 0.537;
0.523; 0.507. Thus, the flow defined by a sample under study has
slowly decreasing correlation. In this situation, it is reasonable
to apply method by Diamond and Alfa \cite{da} oriented to such flows.
 \item[$\bullet$]
 As the result, the process of the arrival of the batches was
 defined by the $MAP$ which is characterized by the matrices
 \begin{gather*}
D_0=\begin{pmatrix}
  -0.0038 & 0 \\
  0 & -0.0066 \\
\end{pmatrix},\;  \\  D_1=A,\;A=\begin{pmatrix}
  0.0037 & 6.58\cdot 10^{-5} \\
  1.3\cdot 10^{-4} & 0.0064 \\
\end{pmatrix}.
\end{gather*}
\item[$\bullet$]
Based on this $MAP$ of batches and information about the batch size
distribution, the $BMAP$ of Web pages is constructed. It is defined
by the matrices $D_0$ and $D_k=d_k A,\; k=\overline{1,8}.$
\end{itemize}

To estimate a Web page service time distribution, the following
information  about the size of an arbitrary information unit
delivered by a crawler (content size) in the used data base was
exploited: the mean content size is equal to 49207.0356 bytes, the
mean squared content size is equal to 2.0527E+11, and the mean cubed
content size is equal to   2.5773E+18. Based on this information,
the service time distribution of a Web page was described, up to
some normalizing constant defined by the content processing rate
(in our application the processing rate is constant), 
by the hyper-exponential distribution which is the partial case of the
$PH$ distribution defined by the vector $ {\boldsymbol \beta} =
(0.0057 \, 0.9943)$ and by sub-generator
$$S=\begin{pmatrix}
  -0.0014  & 0 \\
  0 & -0.2409 \\
\end{pmatrix}.$$ The mean service
time is~8.2. The squared coefficient of variation of the service
time distribution is equal to 86.03. Because the exponential
distribution has the squared coefficient of variation equal to 1, is
is clear that this distribution can not be considered as a good
approximation of the service time distribution.

We assume that the system has four available operation modes. The
buffer capacity is $K=20$.

When the $r$ robots are activated, the $BMAP$-input is described by
the matrices ${\mathcal D}_k^{(r)}$,
$${\mathcal D}_k^{(r)}=rD_k,\quad k=\overline{0,8},r=\overline{1,4},$$
where the matrices $D_k,\quad k=\overline{0,8},$ are defined above.
The intensities $\lambda^{(r)}$ of the $BMAP$ when $r$ robots are
used, $r=\overline{1,4},$ are as follows: $\lambda^{(1)}=0.0153$,
$\lambda^{(2)}=0.0307$, $\lambda^{(3)}=0.046$,
$\lambda^{(4)}=0.061$.

The distribution of a Web page obsolescence time is exponentially
distributed with parameter 0.0005, i.e., the mean time until
obsolescence is equal to 2000.

The cost criterion coefficients are taken as $c_{loss}=200$,
$c_{obs}=250$, $a=3$, $c_{rob}=20$, $c_{star}=600.$

The cost criterion values $C_r$  when $r,$ $r=\overline{1,4},$
robots are always activated are given by $C_1=666.28$, $C_2=657.07$,
$C_3=639.03$ and $C_4=621.25$. When all four modes of operations are
exploited the optimal cost criterion value is 563.51 and the optimal
set of thresholds is [2,2,2], i.e., the optimal strategy assumes
that four robots should be active until the number of Web pages in
the system does not exceed 2. When the number of Web pages in the
system  exceeds 2, three robots should be deactivated and only one
robot should be active.

Table $5$ contains the optimal values of the cost criterion for the
different combinations of operation modes.

\begin{center}
Table $5$: The Values of the Optimal Cost Criterion for the Fixed
Combination of Operation Modes
\begin{tabular}{|c|c|c|}
 \hline possible numbers of active robots & \begin{tabular}{c}
  Optimal \\
  thresholds \\
\end{tabular} & \begin{tabular}{c}
  Optimal value \\
of the cost \\
  criterion \\
\end{tabular}
\\\hline
1&--&666.28\\ 2&--&657.07\\3&--&639.03\\4&--&621.25\\
  1,2&0& 624.97\\
 {1,3} & {0} &{591.72}\\
  \textbf{1,4}&\textbf{2}&\textbf{563.51}\\
  2,3&1&622.81\\
  2,4&2 &593.29 \\
   3,4&3 &609.66 \\
     1,2,3&0,0 &591.72 \\
 1,2,4&2,2 &563.51 \\  1,3,4& 2,2&563.51 \\
2,3,4&2,2 &593.29\\
 1,2,3,4&2,2,2 &563.51 \\
  \hline
  \end{tabular}
\end{center}
The relative profit of operation mode control exceeds 9\% comparing
to the case when no control is applied and all four robots are
always active.

\section{Conclusion}

In this paper we provide performance evaluation and optimization
of the crawling part of a Web search engine. We model the crawler 
with a finite buffer, monotonically controlled arrival rate
(controlled number of crawling robots), and with stochastically 
bounded waiting time. The system is considered under rather general 
assumptions about the arrival process, service and obsolescence 
time distributions. Stationary distribution of the system state, 
sojourn time distribution and main performance measures of the system 
are calculated under any fixed set of thresholds defining the control 
strategy. This allows us to reduce the problem of optimal control 
to minimization of a known function of several integer variables.

Numerical results are presented. They show that the dynamic input
control can give essential profit. Effects of buffer size changes
and changes of average service and obsolescence times and their
variances are investigated. In particular, we illustrate that the
assumption about the exponential distribution of service and
obsolescence times can give poor estimation of the system performance 
measures and optimal values of the thresholds when actually these times 
have high variation.

The model has been applied to the performance evaluation and
optimization of the crawler designed by INRIA Maestro team in the framework
of the RIAM INRIA-Canon research project.


\section*{Acknowledgment}
The authors would like to thank Ministry of Science of France for
financial support of this research via ECO-NET programme and RIAM
INRIA-Canon Research Project.

\onecolumn

\begin{center}
Table 3: Variable Mean Service Time
\\ \vspace{2mm}
\begin{tabular}{|c|c|c|c|c|c|c|c|c|c|}
  \hline
$s$&  $b_1$ &    $\begin{array}{c}
  \text{numbers of} \\
  \text{active robots} \\
\end{array}$ & $j^*$&  $C^*$&    $C_1$&    $C_2$& $C_3$& $C_4$& $R$, \%\\
\hline
0.1&    6.57&   1,3&  0&  53.1&   58.58&  80.67&  104.74& 131.84& 9.35\\
0.2&    3.29&   1,3&  1&  45.08&  59.45&  72.77&  94.16&  122.04& 24.17\\
0.3&    2.19&   1,3&  1&  42.93&  68.51&  72.02&  89.42&  118.55& 37.34\\
0.4&    1.64&   1,3&  1&  43.33&  80.36&  74.27&  86.7&   117.45& 41.66\\
0.5&    1.31&   1,3&  1&  45.28&  93.09&  78.34&  85.15&  117.77& 42.2\\
0.7&    0.94&   1,3&  2&  51.17&  117.98& 89.66&  84.65&  121.17& 39.55\\
0.9&    0.73&   1,3&  2&  58.88&  140.09& 103.05& 87.14&  126.89& 32.43\\
1&  0.66&   1,3&  2&  63.54&  149.91& 110&    89.4&   130.31& 28.93\\
3&  0.22&   1,3&  3&  160.48& 244.04& 211.44& 182.29& 204.54& 11.96\\
5&  0.13&   1,4&  1&  217.02& 272.19& 254.87& 243.01& 251.25& 10.7\\
7&  0.09&   1,4&  1&  250.83& 285.26& 276.93& 274.4&  279.31& 8.59\\
9&  0.07&   1,4&  0&  272.76& 292.74& 290.05& 292.8&  297.61& 5.96\\
11& 0.06&   1,4&  0&  288.23& 297.59& 298.7&  304.77& 310.39& 3.15\\
13& 0.05&   1,4&  0&  299.92& 300.97& 304.82& 313.15& 319.78& 0.35\\
15& 0.04&   1&  --& 309.061&    303.47& 309.37& 319.33& 326.96& 0\\
\hline
\end{tabular}
\end{center}

\newpage
\begin{center}
Table 4: Variable Mean Time until Obsolescence
\\ \vspace{2mm}
\begin{tabular}{|c|c|c|c|c|c|c|c|c|}
  \hline
$s$&  $g_1$ &    $j^*$&  $C^*$&    $C_1$&    $C_2$& $C_3$& $C_4$& $R$, \%\\
\hline 0.01 & 500 & 3 & 52.07 & 130.02 & 92.47 & 81.28 & 118.93 &
37.16 \\
0.1&    50& 3&  52.41&  132.22& 94.24&  81.99&  119.99& 36.08\\
0.2&    25& 2&  53.83&  134.55& 96.17&  82.79&  121.17& 34.98\\
0.3&    16.7&   2&  55.1&   136.78& 98.05&  83.6&   122.34& 34.09\\
0.4&    12.5&   2&  56.36&  138.9&  99.88&  84.41&  123.51& 33.23\\
0.5&    10& 2&  57.6&   140.93& 101.67& 85.24&  124.67& 32.43\\
0.7&    7.1&    2&  60.02&  144.74& 105.12& 86.9&   126.95& 30.93\\
0.9&    5.6&    2&  62.38&  148.25& 108.42& 88.57&  129.21& 29.57\\
1&  5&  2&  63.54&  149.91& 110&    89.4&   130.31& 28.93\\
3&  1.67&   1&  83.6&   173.68& 135.36& 105&    149.91& 20.38\\
5&  1&  1&  95.64&  187.76& 152.48& 117.52& 165.21& 18.62\\
10& 0.5&    1&  116.47& 206.98& 178.08& 138.25& 191.33& 15.75\\
20& 0.25&   0&  131.1&  223.17& 201.46& 158.66& 218.9&  17.37\\
30& 0.17&   0&  136.67& 230.45& 212.45& 168.67& 233.25& 18.97\\
40& 0.13&   0&  139.97& 234.54& 218.83& 174.63& 242.04& 19.85\\
50& 0.1&    0&  142.15& 237.18& 222.99& 178.58& 247.97& 20.4\\
60& 0.083&  0&  143.72& 239.03& 225.91& 181.39& 252.25& 20.77\\
70& 0.071&  0&  144.9&  240.38& 228.08& 183.5&  255.47& 21.04\\
80& 0.063&  0&  145.81& 241.42& 229.75& 185.13& 257.99& 21.24\\
90& 0.056&  0&  146.55& 242.24& 231.08& 186.44& 260.01& 21.4\\
100&    0.05&   0&  147.16& 242.91& 232.15& 187.51& 261.67& 21.52\\
200&    0.025&  0&  150.08& 245.97& 237.19& 192.58& 269.62& 22.07\\
\hline
\end{tabular}
\end{center}]


\end{document}